\newcommand{\diracslash}[1]{#1\llap{/\kern2pt}}

\newcommand{\be}{\begin{equation}}
\newcommand{\ee}{\end{equation}}
\newcommand{\bea}{\begin{eqnarray}}\index{\footnote{}}
\newcommand{\eea}{\end{eqnarray}}
\newcommand{\ba}[1]{\begin{array}{#1}}
\newcommand{\ea}{\end{array}}

\documentclass[12pt]{iopart}
\usepackage{graphicx}
\begin{document}
\setlength{\topmargin}{0.2in}

\title[Magneto-optical Feshbach resonance]{Magneto-optical Feshbach resonance: Controlling cold collision with  quantum interference}
\author{Bimalendu Deb}
\address{Department of Materials Science, Indian Association for the Cultivation of Science,
 Jadavpur, Kolkata 700032. INDIA}

\begin{abstract}
We propose a method of controlling two-atom interaction using both
magnetic and laser fields. We analyse the role of quantum
interference  between  magnetic and optical Feshbach resonances in
controlling cold collision. In particular, we demonstrate that this
method allows us to suppress inelastic and enhance  elastic
scattering cross sections. Quantum interference is shown to modify
significantly the threshold behaviour and resonant interaction of
ultracold atoms. Furthermore, we show that it is possible to
manipulate not only the spherically symmetric s-wave interaction but
also  the anisotropic higher partial-wave interactions  which are
particularly important for high temperature superfluid or
superconducting phases of matter.
\end{abstract}


\maketitle

\section{Introduction}
\label{} Two-particle interaction is a key to describing interacting
many-particle systems at a microscopic level. Means of manipulating
this interaction enable us to explore physics of such systems with
controllable interaction. In solid state systems, the scope of
externally controlling inter-particle interactions is limited due to
crystalline structures.  By contrast, ultracold atomic gases offer a
unique opportunity since their interatomic s-wave interaction is
widely tunable by a magnetic Feshbach resonance (MFR) \cite{1}. New
insight into the exotic phases of interacting electrons in solids
can be gained from the experiments involving ultracold atoms with
tunable interactions. Atom-atom interaction can also be manipulated
by an optical Feshbach resonance (OFR) \cite{2}, albeit with limited
efficiency. Over the last decade, MFR  \cite{revkohler,revgrimm} has
been extensively used to study interacting Bose\cite{3,4,5,5.5} and
Fermi gases\cite{6,7,8} of atoms. Electric fields\cite{9,10}  can
also be used to alter interatomic interaction.

MFR relies on the interplay of Zeeman
effects and hyperfine interactions while OFR is based on
photoassociation (PA)\cite{11,12,13}  of two colliding ground state
atoms into an excited molecular state. OFR has been demonstrated in recent experiments\cite{14,15,16}. Recently, PA spectroscopy in
the presence of an MFR has attracted a lot of attention both
experimentally\cite{17,18,19,bauer:nphys}  and theoretically\cite{20,21,22,23,24}. Junker {\it et
al.}\cite{17} have observed asymmetric profile in PA
spectrum under the influence of an MFR. This spectral asymmetry
results from Fano-type quantum interference\cite{25}  in
continuum-bound transitions\cite{22}. The use of quantum
interference to control Feshbach resonance had been suggested earlier
by Harris\cite{26}. Of late, quantum interference has been observed in
two-photon PA\cite{27,28,29}  and coherent atom-molecule
conversion\cite{30}. It has also been shown that  Fano's
theory\cite{25} can account for PA spectrum\cite{31,32} even in the absence of any MFR.

Here we demonstrate theoretically a new method of altering two-atom
interaction. Let us consider that a laser field is tuned near a PA
transition of two atoms which are simultaneously influenced  by a
magnetic field-induced Feshbach resonance. There are two competing
resonance processes occurring in this system. One is the MFR
attempting to associate the two ground state atoms into a
quasi-bound state embedded in the ground continuum. The other one is
the PA resonance tending  to bind the two atoms into an excited
molecular state.
PA transitions can occur in two competing pathways which originate
from the perturbed and unperturbed continuum states. The Fano-type
quantum interference between these two pathways can be used to
control atom-atom interaction. This quantum control of two-body
interaction due to applied magnetic and optical fields is what we
call ``magneto-optical Feshbach resonance'' (MOFR).  In
strong-coupling regime of PA transitions, s-wave scattering state
gets coupled to higher partial-wave states\cite{33,34}  via
two-photon continuum-bound dipole coupling. Since s-wave scattering
amplitude is largely enhanced due to the applied magnetic field,
amplitudes of the higher partial-waves coupled to s-wave will also
be largely modified. By resorting to a model calculation, we
present explicit analytical expressions for phase shifts, elastic
and inelastic scattering rates which manifestly show the significant
effects of quantum interference in controlling cold collision.
Resonant interaction arises in many physical
situations\cite{35,36,36.1}. It is therefore important to devise
coherent control of resonant interaction.

\section{The model}
\label{}
As a simple model, we consider
three-channel time-independent scattering  of  two homonuclear
Alkali atoms  in the presence of a magnetic and a PA laser field.
Here channel implies asymptotic hyperfine or electronic states of
the two atoms. There are two  ground hyperfine channels of which one
is energetically open (labeled as channel `1') and the other one is
closed (channel `2') in the separated atom limit. Channel 3 belongs
to an excited molecular state which asymptotically corresponds to
two separated atoms with one ground and the other excited atom.  We
assume that the collision energy  is close to the binding energy
of a quasi-bound state supported by the ground closed channel. It is
further assumed that the rotational energy spacing of the excited
molecular levels is much larger than PA laser linewidth so that PA
laser can effectively drives transitions to a single ro-vibrational level ($v, J$) of
the excited molecule, where $v$ stands for vibrational and $J$ for
rotational quantum numbers. The angular state of the two atoms in
the molecular frame of reference can be written as $ \mid J \Omega M
\rangle = i^J \sqrt{\frac{2 J + 1}{8\pi^2}} {\cal D}^{(J)}_{M \Omega
}(\hat{r}) $ where $\Omega$ is the projection of the electronic
angular momentum along the internuclear axis and $M$ is the
z-component of J in the space-fixed coordinate (laboratory) frame.
${\cal D}^{(J)}_{M \Omega }(\hat{r})$ is the rotational matrix
element with $\hat{r}$ representing  the Euler angles for
transformation from body-fixed to space-fixed frame.  In our model,
we assume that the PA laser is tuned near resonance of $J = 1$ level
of the excited molecule.

The energy-normalized dressed state of these three interacting
states with energy eigenvalue $E$ can be written as \bea \Psi_E &=&
\sum_{ M} \frac{\phi_{v J M}(r)}{r} \mid e \rangle \mid J \Omega M
\rangle +
\frac{\chi(r)}{r} \mid g_2 \rangle \mid  0 0 0 \rangle   \nonumber \\
&+& \int d E' \beta_{E'} \sum_{\ell m_{\ell}} \frac{\psi_{ E' \ell
m_{\ell}}(r)}{r} \mid g_1 \rangle \mid \ell 0 m_{\ell} \rangle  \label{dressed} \eea
where $ \phi_{v J M}(r)$ is the radial part of the excited molecular
state, $\chi(r)$ is the bound state in the closed channel and
 $\psi_{ E' \ell m_{\ell}}(r)$ represents energy-normalized scattering
 state of the partial wave $\ell$ with $m_{\ell}$ being the
 projection of $\ell$ along the space-fixed z-axis. $\mid g_i \rangle $
and $\mid e \rangle$ denote the internal electronic states of $i$-th ground
and excited molecular channels, respectively.  Here $E' = \hbar^2 k^2
 /(2\mu)$ is the collision energy, where
$k$ and $\mu$ are the relative momentum and reduced mass of the two
atoms, respectively.   $\beta_{E'}$ denotes density of states of the
unperturbed continuum.
 Note that $ \phi_{v J M}(r)$ and $\chi(r)$ are the perturbed
bound states. In the limit $r \rightarrow \infty$, we have $r \Psi_E
\rightarrow \int d E' \beta_{E'} \sum_{\ell m_{\ell}} \psi_{ E' \ell
m_{\ell}} \mid g_1 \rangle \mid \ell 0 m_{\ell}  \rangle$ and thus
the scattering properties in MOFR are determined by the asymptotic
behavior of $ \psi_{ E' \ell m_{\ell}}$.

From time-independent Schr\"{o}dinger equation,  under
Born-Oppenheimer approximation, we obtain the following coupled
differential equations  \bea &&\left [ \hat{h}_J + V_{e}(r) -
\hbar \delta_1 - E  - i \hbar \gamma_J/2 \right ] \phi_{v J M}
\nonumber \\ &=& - \sum_{\ell,m_{\ell}}\Lambda_{\ell m_{\ell}, J
M}^{(1)} \tilde{\psi}_{E \ell m_{\ell} } + \Lambda_{0 0, J
M}^{(2)}\chi, \label{phi} \eea \bea \left [ \hat{h}_0 + V_{2}(r) - E
\right]\chi = - \sum_{M} \Lambda_{J M, 00}^{(2)} \phi_{v J M}   -
V_{12} \tilde{ \psi}_{E 0 0 } \label{chieq} \eea \bea \left [
\hat{h}_{\ell} + V_{1}(r) - E \right] \tilde{\psi}_{E \ell m_{\ell}
} = &-&  \sum_{M} \Lambda_{\ell m_{\ell}, J
M}^{(1)} \phi_{v J M} \nonumber \\
&-& \delta_{\ell 0} V_{12}  \chi, \label{psi} \eea where
$\tilde{\psi}_{E \ell m_{\ell}}=\int \beta_{E'} d E' \psi_{E' \ell
m_{\ell} }$,   $\hat{h}_{J (\ell)} =
-\frac{\hbar^2}{2\mu}\frac{d^2}{d r^2} + B_{J (\ell)}(r)$ with $B_{J
(\ell)} (r) = \hbar^2/(2\mu r^2) X_{J (\ell)}$ being the rotational
term corresponding to $J (\ell)$. If the excited molecular potential
$V_e$ belongs to Hund's case (a) and (c), then $X_J = [J(J+1) -
\Omega^2]$, otherwise  $X_J = J(J+1)$ and $X_{\ell} = \ell (\ell +
1)$. The laser couplings between different
angular states are denoted by $\Lambda_{\ell m_{\ell}, J M}^{(i)} = -
\langle J M \Omega \mid \vec{D}_i.\vec{{\cal E}}_{PA} \mid \ell m_{\ell} 0 \rangle$,
where $\vec{D}_i$ is the transition dipole moment between the excited and the ground $i$-th channel
molecular electronic states. For homonuclear atoms, $V_{e}(r)$ goes as $- 1/r^3$ and the
ground potentials $V_1$ and $V_2$ behave as $- 1/r^6$ in the limit
$r \rightarrow \infty$. Here $\delta_1 = \omega_1 -  \omega_A $ is
the detuning between the laser  frequency $\omega_1$ and the atomic
resonance frequency $\omega_A$,
 $V_{1(2)}$ is the
interatomic potential in channel 1(2), $\delta_{\ell 0}$ stands for
Kronecker-$\delta$ and  $V_{12}$ denotes  spin-spin coupling between
the two ground channels.  We have here phenomenologically introduced  the term $- i \hbar
\gamma_J/2$ corresponding to the natural linewidth of the excited
state $(v, J)$. The zero of the energy scale is taken to be the threshold of channel 1 and
the atomic frequency $\omega_A$  corresponds to the threshold of the channel 3 (threshold of excited molecular potential). For simplicity, we assume that the excited state
belongs to the $\Sigma$ symmetry. Then the dipole coupling between
angular states provides $m_{\ell} = M$ and thus we can solve the
above coupled equations for a given value of $M$. For notational
convenience, we henceforth suppress the subscripts $M$ and $m_{\ell}$.

\section{The solution}
\label{}

 The coupled equations (\ref{phi}-\ref{psi})  can be
conveniently solved by the method of Green's function.  Let $\phi_{v
J}^{0}$ be the excited bound state solution of the homogeneous part
 of (\ref{phi}) with binding energy $E_{vJ}$.  Using the Green's function  $ G_{vJ}(r,r')= -
\frac{\phi_{J}^{0}(r)\phi_{J}^{0}(r')}{\Delta E_{vJ} + i\hbar
\gamma_J/2 } $ where $\Delta E_{vJ} = \hbar \delta_1 + E - E_{vJ}$,
we can write \be \phi_{v J } (r) = \frac{\int_{E'} d E' \beta_{E'}
\sum_{\ell} \Lambda_{E' \ell,J}   + \Lambda_{bb}}{\Delta E_{vJ} +
i\hbar \gamma_J/2 } \phi_{v J}^{0}(r) \label{phinew1} \ee  where
 $ \Lambda_{E' \ell,J}  = \int dr'\Lambda_{J,
\ell }^{(1)}(r')\phi_{J}^{0}(r') \psi_{E' \ell }(r')$ is the
free-bound dipole coupling between the unperturbed bound state
$\phi_{v J}^{0}$ and the perturbed scattering state  $\psi_{E' \ell
}$ and $ \Lambda_{bb} = \int dr' \Lambda_{J, 0 }^{(2)}(r') (r')
\phi_{v J}^{0}(r') \chi(r')$ is the bound-bound dipole coupling
between $\phi_{v J}^{0}$ and the perturbed bound state $\chi$. Let
$\chi^0(r)$ be the solution of the homogeneous part of (\ref{chieq})
with binding energy $E_{\chi}$. Writing $\phi_{v J }$ in the form
$\phi_{v J } = \int d E'   \beta_{E'} A_{E'} \phi_{v J}^{0}$, we can
express \bea \chi = \frac{1} {E-E_{\chi}} \int   d E' \beta_{E'}
\left ( A_{E'} |\Lambda_{bb}^{0}|^2 +
 V_{E'} \right)
  \chi^0(r) \label{chi} \eea
where $\Lambda_{bb}^{0}$ is the Rabi frequency between the two bound
states $\phi_{v J}^{0}$ and $\chi^{0}$ and  $ V_{E'} = \int dr'
\psi_{E' 0} (r') V_{12} (r') \chi^{0}(r')$. Using this one can
express $\Lambda_{bb}$ in terms of $\Lambda_{bb}^{0}$ and $V_{E'}$.
After having done some minor algebra, we obtain \bea A_{E'} = \frac{(E -
E_{\chi}) \sum_{\ell } \Lambda_{E' \ell,J}^{(1)} + V_{E'}
\Lambda_{bb}^{0}}{(E-E_{\chi}) (\Delta E_{v1} + i \hbar \gamma_J/2)
- |\Lambda_{bb}^{0}|^2 }. \label{ae} \eea Note that the right hand side
of  (\ref{ae}) involves the laser coupling $\Lambda_{E'
\ell,J}^{(1)}$ with the perturbed continuum states. Here $A_{E'}$  is related to the coefficient of $\phi_{v J}^{0}$ in the energy-normalised dressed state (\ref{dressed}) of three interacting states of which two are bound states and one is ground continuum state. Since $\phi_{v J}^{0}$ is unit-normalised, $A_{E'}$ has the dimension of inverse of square root of energy. Physically,  PA excitation probability for collision energies ranging from $E'$ to $E' + d E'$ is given by $|A_{E'}|^2 d E'$.  Now,
substituting (\ref{ae}) into  (\ref{phinew1}) and (\ref{chi}) and then
using the resultant form of $\phi_{v J}$ and $\chi$ into
(\ref{psi}), it is easy to see that the equation of motion for
particular $\ell$-wave function gets coupled to other $\ell$-wave
functions.

The Green's function  for the homogeneous part of (\ref{psi}) can be
written as $ {\cal K}_{\ell} (r,r') =  - \pi\psi_{E
\ell}^{0,reg}(r_<)\psi_{E \ell}^{+}(r_>) $
 where $r_{<(>)}$ implies either $r$ or $r'$ whichever
 is smaller (greater) than the other. Here $\psi_{E \ell}^{+}(r) = \psi_{E \ell}^{0, irr} + i\psi_{E \ell}^{0,reg}$ where $\psi_{E \ell}^{0, reg} $
 and $\psi_{E \ell}^{0, irr}$ represent regular and irregular scattering wave functions, respectively, in the absence of optical and magnetic fields.
  Asymptotically,
  $ \psi_{E \ell}^{0 , reg}(r) \sim
j_{\ell}\cos\eta_{\ell} - n_{\ell}\sin\eta_{\ell}$ and  $ \psi_{E
\ell}^{0, irr}(r) \sim -(n_{\ell}\cos\eta_{\ell} +
j_{\ell}\sin\eta_{\ell}) $, where $j_{\ell}$ and $ n_{\ell}$ are the
spherical Bessel and Neumann functions for partial wave $\ell$  and
$\eta_{\ell}$ is the  phase shift in the absence of laser and
magnetic field couplings. According to Wigner threshold laws, as $k
\rightarrow 0$,  $\eta_{\ell}  \sim k^{2 \ell + 1}$ for $\ell \le
(n-3)/2$, otherwise  $\eta_{\ell} \sim k^{n-2}$ with $n$ being the
exponent of the inverse power-law potential  at large separation.
Using ${\cal K}_{\ell} (r,r')$, the perturbed wave function
$\psi_{E' \ell \ell' }$ can be formally expressed in terms of
$V_{E'}$, $A_{E'}$ and $\Lambda_{bb}^{0}$ and  the partial-wave
free-bound dipole transition matrix elements $\Lambda_{E' \ell, v
J}^{0} = \int d r \phi_J^{0} (r) \Lambda_{\ell,J}^{(1)}(r)
\psi_{E'}^{0}(r) $. Next, substituting this into  (\ref{ae}) and the
expression for $V_{E'}$, we can express $A_{E'}$ exclusively in
terms of couplings between unperturbed states. Explicitly, we have
\bea A_{E'} = \frac{e^{i\eta_0} (q_f + \epsilon)/ (\epsilon + i)
\Lambda_{0} +  \sum_{\ell \ge 1}  e^{i \eta_{\ell}} \Lambda_{E'
\ell, v J}^{0}  }{ {\cal D } - E_{q}^{shift} + i \hbar (\gamma_J +
\Gamma_{q} + \sum_{\ell \ge 1} \Gamma_{J\ell})/2} \label{aefinal}
\eea where $\Lambda_0 = \Lambda_{E' 0, v J}^{(1)}$, $\epsilon = [E -
E_{\chi} - E_{\chi}^{shift}]/(\Gamma_{m f}/2)$ with
 $ E_{\chi}^{shift} =
\rm{Re} \int d r V_{12}(r) \chi^0(r') \int d r'  {\cal K}_{0}(r,r')
V_{12}^*(r')\chi^0(r')$ and $\Gamma_{m f} = 2 \pi \mid \int d r
\psi_{E',0}^{reg,0} (r) V_{12}(r)\chi^0(r)\mid^2 = 2 \pi \mid
V_{E'}^{0} \mid^2$ being  the MFR shift and  line width,
respectively. Here \bea q_f = \frac{V_{eff} + \Lambda_{bb}^{0}}{\pi
\Lambda_{0} V_{E'}^0 } \eea is Fano's $q$-parameter which is, in the
present context, called `Feshbach asymmetry parameter'\cite{22}
with \bea V_{eff} = \rm{Re}  \int d r \phi_{v J}^{0} (r)
\Lambda_{J\ell=0}^{(1)}(r) \int d r' {\cal K}_{0}(r,r')
V_{12}(r')\chi^0(r') \nonumber \eea  being an effective potential
acting between the two bound states as a result of their
interactions with the  s-wave part of the continuum states. In
(\ref{aefinal}),  ${\cal D} = \Delta E_{v J} -  \sum_{\ell} E_{J
\ell}^{shift}$,  $\Gamma_{J \ell} = 2 \pi | \Lambda_{E' \ell, v
J}^{0}|^2$, $E_{J \ell}^{shift} = \rm{Re} \int d r \Lambda_{\ell,
J}^{(1)}(r) \phi_{E' \ell}^0(r') \int d r'  {\cal K}_{\ell}(r,r')
\Lambda_{ J,\ell}^{(1)}(r')\phi_{E'\ell}^0(r') $, \bea \Gamma_q =
\left [\frac{(q_f + \epsilon)^2}{\epsilon^2 + 1} \right ]
\Gamma_{J0} \label{qgamma} \eea
 and
\bea E_{q}^{shift} =  \left [ \frac{\epsilon (q_f^2 - 1) - 2
q_f}{\epsilon^2 + 1} \right ] \frac{\hbar \Gamma_{J0}}{2}.
\label{qshift} \eea  Finally, we have
\bea \psi_{E' \ell } &=& e^{i\eta_{\ell}} \psi_{E'\ell}^{0}
 +    \frac{e^{i \eta_0} V_{E'}^0 + A_{E'}  ( q_f
- i ) \pi \Lambda_{0} V_{E'}^0  }{ (\epsilon + i)\Gamma_f/2 }
\delta_{\ell 0} \nonumber \\
&\times& \int d r'  {\cal K}_{0}(r,r') V_{12}(r')\chi^0(r') + A_{E'}
\int d r' {\cal K_{\ell}}(r,r')
 \Lambda_{ \ell, J}^{(1)} (r') \phi_{v J}^{0}(r') \label{psifinal} \eea
where $\psi_{E\ell}^{0} = \psi_{E \ell}^{0  ,reg}$.
 The equations (\ref{aefinal}) and (\ref{psifinal}) constitute the
solutions of our model.

\begin{figure}
 \includegraphics[width=3.5in]{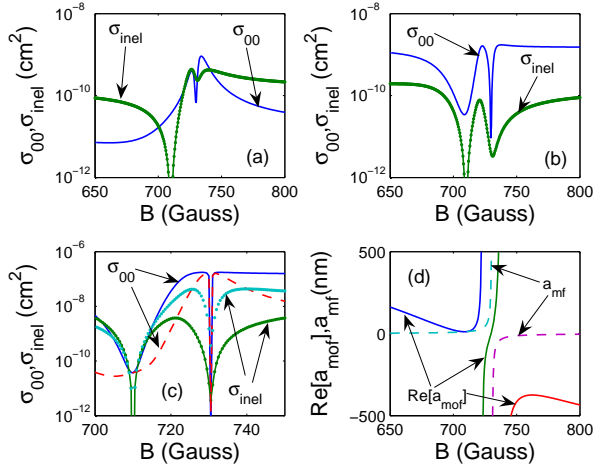}
 \caption{Subplots (a) and (b) show  elastic and inelastic scattering cross
sections $\sigma_{00}$ (solid line) and $\sigma_{inel}$
(solid-dotted line), respectively, in unit of cm$^2$ as a function
of magnetic field $B$ in Gauss (G) for $\Gamma_{J 0}/\gamma=0.1$ (a)
and  $\Gamma_{J 0}/\gamma=10.0$ (b) at collision energy $E=10 \mu$K
and $q_f=-6.89$. Subplot (c) displays $\sigma_{00}$ Vs. $B$ (solid
and dashed lines) and $\sigma_{inel}$ Vs. $B$ (dotted and
solid-dotted lines) plots for
 $\Gamma_{J 0}/\gamma=10.0$ (solid and solid-dotted lines) and  $\Gamma_{J 0}/\gamma=0.1$ (dashed and dotted lines) at $E=100$ nK and $q_f=-68.88$.
Subplot (d) exhibits the variation of Re[$a_{mof}$] (solid line) and
 $a_{mf}$ (dashed lines) as a function of $B$ for $\Gamma_{J 0}/\gamma=10.0$, $E=10 \mu$K
and $q_f=-6.89$. The other fixed parameters for all the subplots are
$\Gamma_{m f} = 16.67$ MHz and $\gamma = 11.7$ MHz.}
 \label{Figure 1.}
  \end{figure}

The  elastic scattering
amplitude is given by $f_{\ell \ell'} = (1/2 i k)( \delta_{\ell
\ell'} - S_{\ell \ell'}) =  T_{\ell \ell'}/k$ where the ${\mathbf
S}$-matrix element $S_{\ell \ell'}$ is related to the  ${\mathbf
T}-$matrix element $T_{\ell \ell'}$ by $S_{\ell \ell'} =
\delta_{\ell \ell'} - 2 i T_{\ell \ell'}$. We can now  derive
$T_{\ell \ell'}$  from the asymptotic behaviour the wave function of
 (\ref{psifinal}) which is given by $\psi_{E' \ell }(r
\rightarrow \infty) \sim  \sin(k r - \ell' \pi /2) \delta_{\ell
\ell'} -  T_{\ell \ell'} \exp(i k r - \ell \pi/2)$. The total elastic scattering
cross section as $\sigma_{el} =  \sum_{\ell',m_{\ell'}} \sum_{\ell, m_{\ell}}
\sigma_{\ell\ell'} $ where $\sigma_{\ell\ell'} = 4 \pi g_s \mid
T_{\ell \ell'} \mid^2/k^2$, with $g_s=1$ for two distinguishable
atoms and $g_s=2$ if the atoms are indistinguishable.

\section{Results and discussions}
\label{}

\subsection{Analytical results}

 We first consider the  s-wave ($\ell = 0$) scattered wave function.
 From the asymptotic form  $\psi_{E',0 } \sim   e^{i\eta_{0}} \psi_{E',0}^{0,reg}  - e^{i (k r +  \eta_0)} [ e^{i \eta_0}
+ A_{E'}  ( q_f + \epsilon ) \pi \Lambda_{0}  ]/( \epsilon + i)  $, we find
 $T_{0 0} = T_0^0 + \exp(2 i \eta_{0}) T_{mf} + \exp[2 i (\eta_{0} + \eta_{mf})]T_q = (1 - S_{00})/2i$ where
$T_0^0 =  - \exp(i \eta_0) \sin \eta_0 $, $T_{mf} =  1/(\epsilon + i) = - \exp(i\eta_{mf}) \sin\eta_{mf}$ where
the MFR phase shift $\eta_{mf}$ is given by
$\cot \eta_{mf} = -\epsilon$,
$ T_q =   \Gamma_q/[{\cal D } - E_{q}^{shift} + i \hbar (\gamma_J + \Gamma_J)]$. Here $\Gamma_J = \sum_{\ell} \Gamma_{J \ell}$. In the limit $k \rightarrow 0$,
$\Gamma_{J \ell} \sim k^{2 \ell + 1}$ and hence $\Gamma_{J 0 } >\!> \Gamma_{J \ell \ne 0}$ for all $\ell \ge 1$. The ${\mathbf S}$-matrix element is $S_{00} = \exp (2 i \eta_{tot})$,
 where
 $\eta_{tot} = \eta_0 + \eta_{mf} + \eta_q$ with
 $\eta_q$  being a complex phase shift. Since in the limit $k \rightarrow 0$, $q_f \sim 1/k $,  near MFR ($\epsilon \simeq 0$) the stimulated linewidth $\Gamma_J \simeq \Gamma_q \simeq q^2
\Gamma_{J 0} \sim 1/k$, $E_q^{shift} \simeq q_f \hbar \Gamma_{J0}$
and  $\cot \eta_q = - [{\cal D}  - E_q^{shift} + i
\gamma_J]/\Gamma_q$. Thus in the limit $\gamma \rightarrow 0$ and $k
\rightarrow 0$,  $T_{ 00}$ fulfills unitarity.

\begin{figure}
 \includegraphics[width=3.5in]{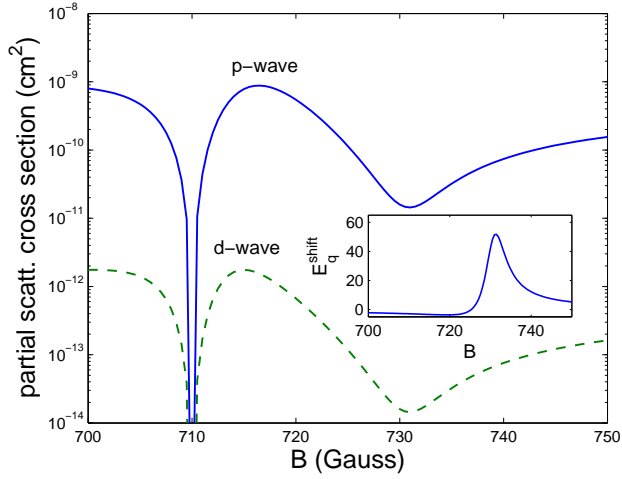}
 \caption{Partial-wave scattering cross section $\sigma_{\ell 0}$ is plotted
as a function of $B$ for $\ell=1$ (solid line) and $\ell=2$ (dashed
lines) for $\Gamma_{J 0}/\gamma=10.0$, $\Gamma_{J 1}=0.1
\Gamma_{J0}$, $\Gamma_{J 2}=10^{-4} \Gamma_{J 0}$, $E=10 \mu$K and
$q_f=-6.89$. The inset shows $E_q^{shift}$ (in unit of $\hbar
\Gamma_{m f}$) as a function of $B$ for the same parameters as in
the main figure. The other
 parameters are same as in figure1 }
 \label{Figure 2.}
  \end{figure}

The s-wave elastic  scattering cross section is $\sigma_{00} = g_s
\pi \mid 1 - S_{00} \mid^2/k^2$ and  the inelastic cross section is
$\sigma_{inel} = g_s \pi (1 - \mid S_{00} \mid^2)/k^2$. The
corresponding rate coefficients are given by $K_{el} = \langle
v_{rel} \sigma_{el} \rangle $ and $K_{inel} = \langle v_{rel}
\sigma_{inel} \rangle $ where $\langle \cdots \rangle $ stands for
thermal averaging over the relative velocity $v_{rel} = \hbar
k/\mu$. Far from MFR ($\epsilon \rightarrow \pm \infty$) we have
$T_{mf} \rightarrow 0$, $E_q^{shift} \rightarrow 0$ and $\Gamma_q
\rightarrow \Gamma_{J0}$. In this limit $T_q$  reduces to the form
$T_{of} =  - \Gamma_{J0}/[{\cal D} + i\hbar(\gamma_J + \sum_{\ell}
\Gamma_{J \ell}]$ which is the ${\mathbf T}$-matrix element of
standard OFR for which  both elastic and inelastic scattering rates
increase as laser intensity increases \cite{37}.

We can define an energy-dependent complex MOFR scattering length by
$a_{mof} = - \tan \eta_{tot}/k $. In the limit $k \rightarrow 0$ we
have \bea a_{mof} \simeq  \frac{ a_{mf} +  q_f^2 \hbar \Gamma_{J
0}/[k({\cal D}  - E_q^{shift} + i \hbar \gamma_J)] } { 1 + k a_{mf}
q_f^2 \hbar \Gamma_{J 0}/({\cal D}
 - E_q^{shift} + i \hbar \gamma_J)}\label{as}\eea
where $a_{mf} = - \lim_{k \rightarrow 0} \tan \eta_{mf}/k$ is the
MFR scattering length. Since $(k  q_f^2 \Gamma_{J 0})$ tends to be
independent of $k$ at ultralow energy,  it is possible to have the
condition Re$[k a_{mf} q_f^2 \hbar \Gamma_{J 0}/({\cal D}
 - E_q^{shift} + i \hbar \gamma_J)] >\!>1$ satisfied  near MFR ($a_{m f} \rightarrow  \pm \infty $) and
PA resonance (${\cal D} \simeq 0$) in the strong-coupling regime ($\Gamma_{J0} >\!> \gamma_J)$.
Note that ${\cal D} = \Delta E_{v J} -  \sum_{\ell} E_{J \ell}^{shift} = 0$
is the PA resonance condition in the absence of MFR. Furthermore, it is to be noted that $E_q^{shift}$ as given by
 (\ref{qshift}) is independent of $k$ in the limit $k \rightarrow 0$ and $\epsilon \rightarrow 0$ and can greatly exceed the
spontaneous linewidth  $\gamma_J$
in the strong-coupling regime\cite{23}.  Under such conditions,
we can write
 \bea a_{mof} \simeq   \left (  \frac{{\cal D}  - E_q^{shift} }
{ k q_f^2 \hbar \Gamma_{J 0}} +  \frac{1}{k^2 a_{mf}} \right ) +
  i \left (\frac{\gamma_J}{ k q_f^2 \Gamma_{J 0}} \right ). \label{ai} \eea
Let us recall that $a_{mf} = - 1/(k \epsilon) = -\hbar
\Gamma_{mf}/[2 k (E' - \tilde{E}_{\chi})]$, where $\tilde{E}_{\chi}
= E_{\chi} + E_{\chi}^{shift}$ and $E'=\hbar^2 k^2/(2\mu)$.
Therefore, in the case of finite  $\tilde{E}_{\chi} > E'$, the real
part of $a_{mof}$ (Re[$a_{mof}$]) becomes inversely proportional to
energy and hence $ \sigma_{el} \sim 1/k^4$ as $k \rightarrow 0$. In the
case of  $\tilde{E}_{\chi}=0$, Re[$a_{mof}$] goes to a constant in
the limit $k \rightarrow 0$. In both the cases, the imaginary part
of $a_{mof}$ (Im$[a_{mof}]$) becomes independent of $k$ but
inversely proportional to laser intensity suggesting that $K_{inel}$
can be made very small by increasing the laser intensity. On the
other hand, for ${\cal D}=0$, the  (\ref{ai}) indicates that
Re$[a_{mof}]$ becomes independent of laser intensity. Thus we can
infer that
 the inelastic scattering rate can be suppressed
while elastic rate can be enhanced by using quantum interference in
the strong-coupling regime at ultralow temperatures. Very recently, Bauer {\it et al.} \cite{bauer:nphys,bauer:pra} have experimentally demonstrated the effect of 
suppression of inelastic rate in PA due to the influence of a magnetic Feshbach resonance.

The amplitudes of higher partial-wave scattered wavefunctions can
also be enhanced by MOFR. The  higher partial waves that can be
manipulated are given by the condition $\vec{J} = \vec{L} + \vec{S}
+ \vec{\ell}$. In the case of singlet to singlet PA transition for
$J=1$, the maximum partial-wave that can be significantly affected
is $\ell = 2$ (d-wave), while in the case of triplet to triplet
transition it is $\ell=3$. For $\ell \ne 0$, we have $ T_{\ell 0} =
\pi  A_{E'} \exp(i \eta_{\ell})\Lambda_{E \ell, v J}^{0} $. Using
(\ref{aefinal}),  in the leading order in dipole coupling at
ultralow energy we have  \bea T_{\ell 0} \simeq \frac{ e^{i (\eta_0
+ \eta_{\ell})} (q_f + \epsilon)/ (\epsilon + i) \pi \Lambda_{0}
\Lambda_{E \ell, v J}^{0}}{ {\cal D } - E_{q}^{shift} + i \hbar
(\gamma_J + \Gamma_{q})/2} \eea
 In the limit $\epsilon \rightarrow \infty$, $T_{\ell,0}$ reduces to that
of OFR\cite{33} for $\ell \ge 1$

\begin{figure}
 \includegraphics[width=3.5in]{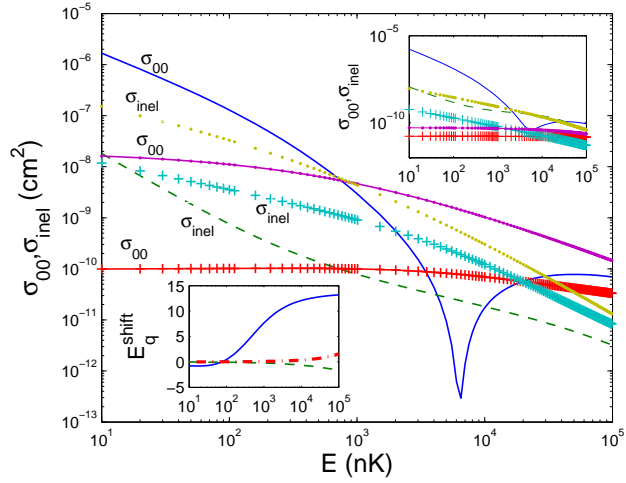}
 \caption{$\sigma_{00}$ and $\sigma_{inel}$  are plotted as a function of
collision energy $E$ (in nK) for $B=730$ G (solid and dashed
curves), $B=700$ G (plus-solid and plus curves) and $B=800$ G
(solid-dotted and dotted curves) with $\Gamma_{J 0}/\Gamma_{m f} =
1.4$. The upper inset shows the same but for $\Gamma_{J 0}/\Gamma_{m
f} =0.07$. In the lower inset, $E_q^{shift}$ (in unit of $\hbar
\gamma$) is plotted against $E$ for $B=730$ G (solid line), $B=700$
G (dotted line) and $B=800$ G (dashed lines) with $\Gamma_{J
0}/\Gamma_{m f} = 1.4$. The other parameters remain same as in
figure1}
 \label{Figure 3.}
  \end{figure}

\subsection{Numerical results}

To illustrate further the analytical results discussed above, we
present selective numerical results. As a model system, we consider
$^7$Li atoms with PA transition $ {^3}\Sigma^+_u \rightarrow
{^3}\Sigma_g^{+}$ . The
parameter $\epsilon$ is related\cite{38}  to the magnetic field
$B$, the resonance  width $\Delta$ and the background scattering
length  $ a_{bg}$ by $\epsilon \simeq - (B - B_0)/(  k a_{bg}
\Delta)$,
 where $B_0$ is the  resonance magnetic field.
We use the realistic parameters taken or estimated from earlier
experimental results\cite{39,40}. These parameters are the
spontaneous line width $\gamma_{J} = 11.7$ MHz \cite{39},
$\Delta = - 192.3$ Gauss (G) and  $ a_{bg} =-24.5 a_0$ ($a_0$ is
Bohr radius). We take $B_0 = 730.5$ G. From the reported Fano
profile of PA spectrum\cite{17}, we extract $q_f = -
6.89$ at $E = 10 \mu$K. Using low energy behaviour $q_f \sim 1/k$,
we extrapolate $q_f$ at other collision energies.  The Feshbach
resonance line width $\Gamma_{mf}$ is taken to be 16.66 MHz   for
$E=10 \mu$K. In all our numerical plots we set ${\cal D}=0$.

In figure 1 (a-c), $\sigma_{00}$ as a function of $B$ is compared
with $\sigma_{inel}$. We notice that, compared to weak-coupling
results of figure 1(a), the strong-coupling result
 $\sigma_{00}$ in figure 1(b) largely exceeds $\sigma_{inel}$ in almost
entire range of $B$. Because of interference between the two
resonances, two closely spaced maxima appears near $B_0$ in figure
1(b). Even in figure 1(a), there is a prominent maximum at and near
which $\sigma_{00}$ exceeds $\sigma_{inel}$. The reason for such
feature is that, as can be inferred from   (\ref{ai}),  for a given
collision energy and ${\cal D}=0$, Re$[a_{mof}]$ becomes independent
of laser intensity as $\epsilon \rightarrow 0$ while Im$[a_{mof}]$
goes to zero in the strong-coupling regime. Figure 1(c) shows that
at much lower energy ($E=100$ nK) inelastic scattering rates are
further suppressed while elastic ones are enhanced both in weak- and
strong-coupling regimes. figure 1(d) illustrates how MFR is split
into a double-resonance owing to Fano interference. This explains
the appearance of two peaks  near $B_0$. The minimum at $B=710$ G
arises due to Fano minimum at which PA transition amplitude
vanishes.

We show the partial p- and d-wave scattering amplitudes in figure 2
in the strong coupling regime. Typically, the higher partial-wave
stimulated line width $\Gamma_{J \ell=1}$ and $\Gamma_{J \ell=2}$
are smaller than $\Gamma_{J \ell=0}$ by one and four order of
magnitudes, respectively\cite{33}.  Comparing figure 2 with figure
1(b), we notice that p- and d-wave scattering cross sections show a
maximum near $B_0$ at which $\sigma_{\ell=1,0}$ is of the same order
of $\sigma_{00}$ while $\Gamma_{J \ell=2}$ is 3 order of magnitude
smaller than that $\sigma_{00}$. The minimum near $B \simeq 730$ G
can be attributed to the quantum interference induced anomalously
large positive shift as shown in the inset of figure 2.

Figure 3 shows energy dependence of elastic and inelastic scattering
cross sections at three different values of $B$ in both the strong-
(main figure) and weak-coupling (upper inset) regimes. The main figure and the upper inset clearly
show  that when $B=730$ G which is close to $B_0$, the elastic part of scattering cross
section largely exceeds the inelastic part in the low energy regime. We notice that elastic scattering cross section $\sigma_{00}$ (solid curve) at $E=10$ nK and $B=730$ G exceeds the inelastic scattering cross section $\sigma_{inel}$ (dashed curve) by two orders of magnitudes. In contrast, this does not happen if $B$ is tuned far away from
$B_0$. For instance, when $B=700$ G and $E=10$ nK, $\sigma_{00}$ (plus solid curve) is smaller than $\sigma_{inel}$ (plus curve) by two orders of magnitude. The effect of laser intensity on the scattering cross sections at low energy can be understood by comparing the main figure with the upper inset of figure 3. The stimulated line width ($\Gamma_{J0}$) in the strong-coupling regime (main figure) is taken to be twenty times larger than that in the weak-coupling regime (upper inset). In other words, PA laser intensity for strong-coupling case is taken to be twenty times larger compared to the weak-coupling case. Let us now compare the plots of
the main figure with the corresponding plots of the upper inset: When $B$  is tuned close to $B_0$ or MFR,  the elastic scattering cross section  $\sigma_{00}$ (solid curve) for strong- (main figure) as well as weak-coupling (upper inset)  regime tends to be equal as the energy $E$ decreases. At $E=10$ nK, we find $\sigma_{00} \simeq 1.7 \times 10^{-6}$ cm$^2$ in both the regimes. In contrast, when $B=700$ G which is away from MFR, $\sigma_{00}$ (plus solid curves) at $E=10$ nK for weak- and strong-coupling regimes are $1.7 \times 10^{-11}$ cm$^2$ and $9.9 \times 10^{-11}$ cm$^2$, respectively. Thus in conformity with our previous analysis,
by comparing the plots in the main and in the upper inset of figure 3,  we can infer that when $B$ is tuned near $B_0$, the elastic cross section at low energy becomes independent of laser intensity.
The minimum at $B \simeq B_0$  in $\sigma_{00}$ Vs. $E$ plots of  figure 3 can be attributed to the large
positive shift $E_q^{shift}$ as depicted in the lower inset of this figure.

\section{Conclusions and outlook}
Quantum interference is shown to change threshold and resonance
behviour significantly. This may in turn change the character of
near-zero energy dimer states. Therefore, the crossover physics
between Bardeen-Cooper-Schrieffer (BCS) state of atoms and
Bose-Einstein condensate (BEC) of such dimers are likely to be
affected by MOFR. Although MFR can most efficiently tune s-wave
scattering length, there exists no standard method of tuning higher
partial-wave interatomic interaction.  MOFR will  be particularly
useful for tuning higher partial-wave interaction. MFR is not
applicable for atoms having no spin magnetic moment and so is MOFR.
However, the underlying principle of MOFR can also be applicable to
such atoms provided a quasi-bound state embedded in the ground
continuum is tunable by a nonmagnetic means.

\appendix
\section*{Appendix-A}
\setcounter{section}{1}
 We discuss how to derive  (8).  Using
${\cal K}_{\ell}$   we first convert  (4) (with the index $M =
m_{\ell}$ being suppressed) into an integral equation of the form
\bea \tilde{\psi}_{E \ell }( r ) &=&
\exp({i\eta_{\ell}})\tilde{\psi}_{E\ell}^{0} +
 \int d r' {\cal K}_{\ell}(r,r') \nonumber \\
&\times& \left [ \Lambda_{ \ell, J}^{(1)} (r') \phi_{v J} (r')  +
V_{12}(r')  \chi(r') \delta_{\ell 0} \right ]  \label{A1} \eea
Substituting $\phi_{v J } = \int d E' \beta_{E'} A_{E'} \phi_{v
J}^{0}$ and  (6)  into  (\ref{A1}), we get \bea \psi_{E' \ell
} &=& e^{i\eta_{\ell}} \psi_{E'\ell}^{0} +
\frac{\Lambda_{bb}^{0} A_{E'} + V_{E'} }{E-E_{\chi} } \delta_{\ell
0} \nonumber \\ &\times& \int d r' {\cal K}_{0}(r,r') V_{12}
(r')\chi^0(r')  \nonumber \\ &+&  A_{E'} \int d r' {\cal
K_{\ell}}(r,r')
 \Lambda_{ J  \ell }^{(1)}(r') \phi_{v J}^{0}(r')
  \label{A2}
\eea
Putting the above equation for $\ell=0$ ($\psi_{E' 0 }$) into the equation $V_{E'} = \int d r \psi_{E' 0 }(r) V_{12}(r) \chi^0(r)$ and after a minor algebra  we obtain
\bea
V_{E'} &=&   \frac{\left ( E - E_{\chi} \right ) \left [e^{i \eta_0} V_{E'}^0
+ A_{E'} \left ( V_{eff} - i \pi  \Lambda_0 V_{E'}^0 \right ) \right ]}{E - (E_{\chi} + E_{\chi}^{shift}) + i\Gamma_f/2} \nonumber \\
&+& \frac{ A_{E'} \Lambda_{bb}^{0} \left ( E_{\chi}^{shift} - i
\Gamma_f/2 \right )}{E - (E_{\chi} + E_{\chi}^{shift}) + i \Gamma_{m
f}/2}  \label{A3}
 \eea
After having substituted  (\ref{A3}) into  (\ref{A2}), we are left
with the only unknown  parameter $A_{E'}$. Now, substituting
(\ref{A2}) and (\ref{A3}) into  (7) and using  $\epsilon = [ E -
(E_{\chi} + E_{\chi}^{shift} )]/(\Gamma_{m f}/2)$ and the parameter
$q_f$ defined by  (9), we obtain  (8).  Thus   (12) is finally
expressed in terms of all the known or unperturbed parameters.

\end{document}